\documentclass[useAMS,usenatbib]{mn2e}
\usepackage{graphicx}
\usepackage{amssymb}
\usepackage{color}

\newcommand{\xmm}{\textit{XMM-Newton}}

\title[Non-thermal emission from old supernova remnants]{
Non-thermal emission from old supernova remnants}
\author[Jun Fang, Li Zhang]{Jun Fang$^1$\thanks{email: fangjun1653@126.com}, Li Zhang$^{1,2}$ \\
$^{1}$Department of Physics, Yunnan University, Kunming, PRC\\
$^{2}$National Astronomical Observatories/Yunnan Observatory,
Chinese Academy of Sciences, P.O. Box 110, Kunming, PRC}

\begin{document}
\pagerange{\pageref{firstpage}--\pageref{lastpage}} \pubyear{2007}

\maketitle

\label{firstpage}

\begin{abstract}
We study the non-thermal emission from old shell-type supernova
remnants (SNRs) on the frame of a time-dependent model. In this
model, the time-dependent non-thermal spectra of both primary
electrons and protons as well as secondary electron/positron
($e^{\pm}$) pairs can be calculated numerically by taking into
account the evolution of the secondary $e^{\pm}$ pairs produced from
proton-proton (p-p) interactions due to the accelerated protons
collide with the ambient matter in an SNR. The multi-wavelength
photon spectrum for a given SNR can be produced through leptonic
processes such as electron/positron synchrotron radiation,
bremsstrahlung and inverse Compton scattering as well as hadronic
interaction. Our results indicate that the non-thermal emission of
the secondary $e^{\pm}$ pairs is becoming more and more prominent
when the SNR ages in the radiative phase because the source of the
primary electrons has been cut off and the electron synchrotron
energy loss is significant for a radiative SNR, whereas the
secondary $e^{\pm}$ pairs can be produced continuously for a long
time in the phase due to the large energy loss time for the p-p
interaction. We apply the model to two old SNRs, G8.7$-$0.1 and
G23.3$-$0.3, and the predicted results can explain the observed
multi-wavelength photon spectra for the two sources.
\end{abstract}


\section{Introduction} \label{sec:intro}

In recent years, various observational evidences indicate that
particles in SNRs can be accelerated up to relativistic energies.
For example, X-ray observations show that electrons can be
accelerated to about 10 TeV in the shell of an SNR (Koyama et al.
1995; Reynolds 1996; Bamba et al. 2003, 2005a, 2005b). TeV
$\gamma$-ray observations indicate that both electrons and protons
can be accelerated to several tens TeV in young shell-type SNRs such
as RX~J1713.7$-$3946 (Enomoto et al. 2002; Aharonian et al. 2006a)
and RX~J0852.0$-$4622 (Katagiri et al. 2005; Aharonian et al. 2005).
More recently, a survey of the inner part of the Galactic Plane in
VHE $\gamma$-ray band has been performed with \textit{HESS}
Cherenkov telescope system. Aharonian et al. (2006b) presented
detailed spectral and morphological information of some new sources
discovered in the \textit{HESS} survey along with some discussions
on possible counterparts in other wavelength bands. Some sources
discovered in the survey can be associated with old SNRs. For
example, HESS J1804$-$216 has a flux of nearly 25\% of the flux of
the Crab Nebula with a photon index of about 2.72. The source can be
associated with either the shell-type SNR G8.7$-$2.1 or the young
Vela-like pulsar PSR J1803$-$2137 (Aharonian et al. 2006b).
Furthermore, Fatuzzo, Melia, \& Crocker (2006) found that the SNR
G8.7$-$2.1 can plausibly account for all known radiative
characteristics of HESS J1804$-$216. Another case is
HESS~J1834$-$087, which is an extended TeV source discovered in the
\textit{HESS} survey of the inner Galaxy in VHE $\gamma$-ray band.
The source, spatially coincident with the SNR G23.3$-$0.3, has a
$\gamma$-ray flux of 8\% of the flux of the Crab Nebula above 200
GeV, and the $\gamma$-ray spectrum can be expressed by a power law
with a photon index of $2.45\pm0.16$ (Aharonian et al. 2006b). The
new \xmm\ observation reveals diffuse X-ray emission within the
\textit{HESS} source and suggests an association between the diffuse
X-rays and the VHE $\gamma$-rays, moreover, G23.3$-$0.3 is estimated
to be an old SNR with a distance of $4\pm0.2$ kpc and an age of
$\sim10^5$ yr (Tian et al. 2007).

Theoretically, it is believed that the non-thermal photon spectrum
for a given SNR is produced through electron/positron synchrotron
radiation, bremsstrahlung and inverse Compton scattering as well as
$\pi^0$ decay from the p-p interactions. The question is, however,
whether the origin of the non-thermal photons for young SNRs is
different from that for old ones. Sturner et al. (1997) presented a
temporal evolution model to calculate the non-thermal particle and
photon spectra from shell-type SNRs, and Zhang \& Fang (2007)
modified the model to investigate the possible hadronic
contributions to TeV $\gamma$-ray emission from young SNRs. For the
young SNRs, the secondary particles produced via the p-p interaction
have a negligible contribution to the non-thermal photon spectrum
(e.g. Zhang \& Fang 2007). On the other hand, Yamazaki et al. (2006)
studied the emission from old SNRs and found that proton
acceleration could be efficient enough to emit TeV $\gamma$-rays
both at the shock of the SNR and at the giant molecular cloud. They
presented that the energy flux ratio $\rm{R}_{\rm{TeV/X}}$ can be
more than $10^2$ for some SNRs, and such sources may be the origins
of the recently discovered unidentified TeV sources. For the old
SNRs, the energies of the primary electrons accelerated by the shock
wave are limited by the synchrotron cooling, then the roll-off
energy of the synchrotron emission of the primary electrons is much
smaller than that for the young SNRs (Sturner et al. 1997; Yamazaki
et al. 2006; Zhang \& Fang 2007). However, the secondary $e^{\pm}$
pairs can be produced from the p-p collisions in an SNR for a long
time due to the large energy loss time for the p-p interaction.
Therefore, the contributions of secondary electrons and positrons to
the non-thermal emission can become prominent for the old SNRs.

Motivated by above discussions, we study the non-thermal photon
spectra from old SNRs based on the time-dependent model in which the
contribution of the secondary $e^{\pm}$ pairs is included (see \S
2.2). It should be noted that our model is different from that given
by Yamazaki et al. (2006), the latter assumes a steady particle
distribution. This paper is organized as follows. The temporally
evolving model of the non-thermal particle and photon spectra are
discussed in Section \ref{sec:model}, including the SNR evolution,
shock acceleration, and various photon production mechanisms
involved in the model. The results from the applications of the
model to two old SNRs, G8.7$-$0.1 and G23.3$-$0.3, are shown in
Section \ref{sec:application}. The main conclusions and some
discussions are given in Section \ref{sec:discussion}.
\section{Temporal Model for Multi-wavelength Non-thermal Emission from SNRs} \label{sec:model}

\citet{S97} presented a temporally-evolving model for the
non-thermal particle and photon spectra at different stages in the
lifetime of a shell-type SNR. Zhang \& Fang (2007) modified the
model to explain the multi-wavelength observations of the young SNRs
and shown that the TeV $\gamma$-rays from the two young shell-type
SNRs RX~J1713.7$-$3946 and RX~J0852.0$-4622$ have hadronic origin.
In this paper, we include the evolution of secondary $e^{\pm}$ pairs
to model the multi-wavelength non-thermal emission from old SNRs
since the secondary $e^{\pm}$ pairs can have significant
contributions to the non-thermal emission from old SNRs.

\subsection{SNR Evolution and Shock Acceleration}

In the analytical model of the shock dynamics of an SNR expanding
into a uniform ambient medium with density $n_0$, the SNR evolves
through three stages: the free expansion stage, the Sedov stage, and
the radiative stage (Sturner et al. 1997; Yamazaki et al. 2006).
Assuming the initial explosion energy of the SNR is
$E=10^{51}E_{51}$ ergs, and the initial velocity of the shock is
$v_0$, the free expansion stage ends when $t=t_{\rm{Sed}}=(3E/2\pi
m_{\rm{H}}n_0v^5_0)^{1/3}\approx 2.1\times
10^2(E_{51}/n_0)^{1/3}v^{-5/3}_0$ yr, and the Sedov stage ends at
$t=t_{\rm{rad}}\approx 4.0\times 10^4E^{4/17}_{51}n^{-9/17}_0$ yr
(Blondin et al. 1998), where $n_0 = \mu n_{\rm{ISM}}$,
$n_{\rm{ISM}}$ is the hydrogen density in the local interstellar
medium, $\mu=1.4$ is the mean atomic weight of the interstellar
medium assuming 1 helium atom for every 10 hydrogen atoms, and
$m_{\rm{H}}$ is the mass of hydrogen. The shock velocity
$v_{\rm{s}}(t)$ can be expressed as (e.g. Yamazaki et al. 2006)
\begin{equation}
v_{\rm{S}}(t)=\left\{ \begin{array}{lll}
v_0                                   &  t<t_{\rm{Sed}} \\
v_0\left(\frac{t}{t_{\rm{Sed}}}\right)^{-3/5}     &  t_{\rm{Sed}}\leq t <t_{\rm{rad}} \\
v_0\left(\frac{t_{\rm{rad}}}{t_{\rm{Sed}}}\right)^{-3/5}
\left(\frac{t}{t_{\rm{rad}}}\right)^{-2/3} &  t_{\rm{rad}} \leq t\;.
\end{array} \right.
\label{vst}
\end{equation}
The evolution of the shock radius is estimated by
$R_{\rm{s}}(t)=\int v_{\rm{s}}(t)dt$ and we assume
$v_{0}=10^{9}$cm~s$^{-1}$ in this paper. It should be noted that the
accumulated energy in non-thermal protons for an SNR is usually
smaller than $10\%$ of the initial explosion energy (see details in
\S 2.1), and the energy diminishes very slowly in the radiative
phase when the thermal energy in the SNR experiences significant
loss. At late times, the relativistic protons begin to contribute
significantly to the post-shock pressure when most thermal energy of
the SNR is drained and then the expansion law should be revised
correspondingly. However, the expansion law will be influenced
significantly by the accelerated protons at very late times and the
non-thermal emission from old SNRs does not depend heavily on the
expansion law. Therefore, we still assume the expansion law
expressed by Eq. (\ref{vst}) is always valid in this paper.

It is generally believed that the non-thermal charged particles are
produced through diffusive shock acceleration (Blandford \& Eichler
1987; Ellison, Jones \& Reynolds 1990). Neglecting the nonlinear
effects and assuming a compression ratio near 4, the particle
spectrum from the diffusive acceleration process is a power law with
spectral index $\alpha\sim 2$. For the particles accelerated in the
shock of an SNR, the power law particle spectra do not extend to
infinite energy and must be truncated by three mechanisms, namely,
the finite age of the SNR, energy-loss processes, and free escape
from the shock region (Sturner et al. 1997). The maximum kinetic
energy of electrons and protons, $E_{e,~\rm{max}}$ and
$E_{p,~\rm{max}}$, can be calculated by using Eqs. (3), (4), (5) in
Sturner et al. (1997), and we neglect the exact formulae here. Zhang
\& Fang (2007) studied the non-thermal emission from three young
SNRs, and concluded that the model in the paper can explain the
multi-wavelength observations for the three young SNRs with the
maximum wavelength of MHD turbulence $\lambda_{\rm{max}}=2\times
10^{17}$cm and other appropriate parameters. $\lambda_{\rm{max}}$ is
also set to $2\times 10^{17}$cm in this paper.

 It should be pointed out that a time-dependent kinetic equation
could be solved in order to obtain the energy spectrum of the
accelerated particles (Berezhko et al. 1996; Sturner at al. 1997).
Here, following Sturner at al. (1997), we approximate the
volume-averaged emissivity, $Q(E,t)=dN/dVdtdE$, of the shock
accelerated electrons and protons by
\begin{eqnarray}
\nonumber Q_{e}^{\rm{pri}}(E,t) &=&
Q^0_{e}G(t)[E(E+2m_ec^2)]^{-[(\alpha+1)/2]}\\
& & \times(E+m_ec^2)\exp(-\frac{E}{E_{e,~\rm{max}}(t)})\;\;,
\end{eqnarray}
and
\begin{eqnarray}
\nonumber Q_{p}^{\rm{pri}}(E,t) &=&
Q^0_{p}G(t)[E(E+2m_pc^2)]^{-[(\alpha+1)/2]}\\
& & \times(E+m_pc^2)\exp(-\frac{E}{E_{p,~\rm{max}}(t)})\;\;,
\end{eqnarray}
respectively, where $E$ is the particle kinetic energy, and $m_e$
and $m_p$ are electron mass and proton mass, the index
$\alpha\sim2.0$, and we use $\alpha=2.0$ in this paper. The function
$G(t)$ is given by (Sturner at al. 1997)
\begin{equation}
G(t)=\left\{ \begin{array}{ll}
\left[R_{\rm{SNR}}(t_{\rm{Sed}})/R_{\rm{SNR}}(t)\right]    &  t\leq t_{\rm{rad}} \\
0 &  t>t_{\rm{rad}}\;.
\end{array} \right.
\label{gt}
\end{equation}
Factors $Q^0_e$ and $Q^0_p$ are used to normalize the particle
spectra so that the total amount of the kinetic energy contained in
the injected electrons and protons is $E_{\rm{par}}=\eta
M_{\rm{ej}}v^2_0/2$, where $\eta\sim0.1$ represents the efficiency
that the kinetic energy of the ejecta with initial mass
$M_{\rm{ej}}$ and initial velocity $v_0$ is converted into the
kinetic energy of both the electrons and the protons,
\begin{eqnarray}
\nonumber E_{\rm{par}}&=&
\int^{t_{\rm{rad}}}_0dtV_{\rm{SNR}}(t)(\int^{E_{e,~\rm{max}}}_0dE
EQ^{\rm{pri}}_{e}(E,t)\\
& & {}+\int^{E_{p,~\rm{max}}}_0dE EQ^{\rm{pri}}_{p}(E,t))\;\;,
\label{Etot}
\end{eqnarray}
and $V_{\rm{SNR}}(t)=4\pi R^3_{\rm{SNR}}(t)/3$ is the SNR volume. In
Eq. (\ref{Etot}), we need to introduce the parameter
$K_{\rm{ep}}=Q^0_e/Q^0_p$ in order to determine  $Q^0_e$ and
$Q^0_p$. Obviously, these two quantities depend on $\eta$ and
$K_{\rm{ep}}$. The ratio of electrons to protons, $K_{\rm{ep}}$, is
a key parameter when modeling the multi-wavelength non-thermal
emission from an SNR. The measured ratio from the cosmic ray
observations at the Earth is about 0.013 at 10 GeV (Gassier 1990).
However, the measured value can not represent the ratio for a given
SNR due to the cosmic ray transport effect in the Galaxy. Sturner et
al. (1997) obtained $K_{\rm{ep}}\sim0.6$ by assuming that the
kinetic energy of the injected electrons is the same as that of the
injected protons. Yamazaki et al. (2006) studied the TeV emission
from the old SNRs in the steady state, they used the observed ratio
of TeV $\gamma$-ray (1--10 TeV) to X-ray (2--10 keV) energy flux,
$R_{\rm{TeV/X}}$, for some SNRs to test the values of $K_{\rm{ep}}$,
and found that the predicted $R_{\rm{TeV/X}}$ for young SNRs is
similar to the observed one when $K_{\rm{ep}}=1\times10^{-3}$, and
that the predicted values of $R_{\rm{TeV/X}}$ for the old SNRs are
not changed even if $K_{\rm{ep}}$ varies more than one order of
magnitude because the primary electrons do not contribute to TeV and
X-ray emission in their model. Moreover, Berezhko, Ksenofontov \&
V$\ddot{\rm{o}}$lk (2006) investigated the properties of Kepler's
SNR using the nonlinear kinetic theory of cosmic ray acceleration in
the SNR. They treated $K_{\rm{ep}}$ as a parameter quantitatively
determined by comparison with the synchrotron observations and
obtained that $K_{\rm{ep}}\sim 10^{-4}$ for the SNR. In fact, the
value of $K_{\rm{ep}}$ for a given source is usually uncertain now,
and it can be limited by comparison of the resulting spectrum with
the multi-wavelength observations for an SNR (Zhang \& Fang 2007).
In this paper, $\eta$ is set to 0.1 and $K_{\rm{ep}}$ is treated as
a parameter limited by comparison of the calculated results with the
observations for a given SNR.

\subsection{Temporal Evolution of Particle Energy Distributions}

Following Sturner et al. (1997), assuming that an SNR interior is
homogeneous, with a constant density $n_{\rm{SNR}}=4n_{\rm{ISM}}$
and a magnetic field strength $B_{\rm{SNR}}=4B_{\rm{ISM}}$,
$n_e(E_e,t)$ and $n_p(E_e,t)$ are used to represent the differential
densities of the accelerated electrons and protons, respectively. We
can use $Q_e$ and $Q_p$ to calculate the direction- and
volume-averaged electron intensity
$J_e(E_e,t)=(c\beta/4\pi)n_e(E_e,t)$ and proton intensity
$J_p(E_p,t)=(c\beta/4\pi)n_p(E_p,t)$ at each moment during the SNR
lifetime. Fokker-Planck equations for both electrons and protons in
energy space are used to find solutions, which are given by (Sturner
et al. 1997)
\begin{eqnarray}
\nonumber \frac{\partial n_e(E_e,t)}{\partial t}&=&-\frac{\partial
}{\partial
E_e}\left[\dot{E}^{\rm{tot}}_en_e(E_e,t)\right]\\\nonumber & &
{}+\frac{1}{2}\frac{\partial^2}{\partial
E^2_e}\left[D(E_e,t)n_e(E_e,t)\right]\\
& & {} +Q_e(E_e,t) -\frac{n_e(E_e,t)}{\tau}\;\;, \label{Net}
\end{eqnarray}
and
\begin{eqnarray}
\nonumber \frac{\partial n_p(E_p,t)}{\partial t}&=&-\frac{\partial
}{\partial
E_p}\left[\dot{E}^{\rm{tot}}_pn_p(E_p,t)\right]\\
\nonumber & & {}+\frac{1}{2}\frac{\partial^2}{\partial
E^2_p}\left[D(E_p,t)n_p(E_p,t)\right]\\
& & {}+Q_p(E_p,t)-\frac{n_p(E_p,t)}{\tau}\;\;, \label{Npt}
\end{eqnarray}
respectively, where the terms on the right-hand sides in Eqs.
(\ref{Net}) and (\ref{Npt}) represent systematic energy losses,
diffusion in energy space, the particle source function and
catastrophic energy loss. Following Sturner et al. (1997), we solve
the above equations using a Crank-Nicholson finite difference
scheme. In Eqs. (\ref{Net}) and (\ref{Npt}), $\dot{E}^{\rm{tot}}_e$,
$D(E_e,t)$, $\dot{E}^{\rm{tot}}_p$ and $D(E_p,t)$ can be calculated
with the formulae in \S 2.2 in Zhang \& Fang (2007), the exact
expressions are neglected in this paper.

The secondary electrons and positrons can be produced when the
accelerated protons collide with the ambient matter via the p-p
interaction. Those secondary particles evolve with the SNR, and can
contribute to the multi-waveband non-thermal emission from the SNR
as the primary electrons. So the source term for electron can be
represented as
\begin{equation}
Q_e(E,t)=Q_e^{\rm{pri}}+Q_{e^+}^{\rm{sec}}(E,t)+Q_{e^-}^{\rm{sec}}(E,t)\;\;,
\label{Qetot}
\end{equation}
where
\begin{equation}
Q_{e^+}^{\rm{sec}}(E,t)=4\pi \mu_{\rm{pp}} n_{\rm{SNR}}\int dE_p
J_p(E_p,t)\frac{d\sigma(E_{e^+},E_p)}{dE_{e^+}}, \label{Qe+}
\end{equation}
\begin{equation}
Q_{e^-}^{\rm{sec}}(E,t)=4\pi \mu_{\rm{pp}} n_{\rm{SNR}}\int dE_p
J_p(E_p,t)\frac{d\sigma(E_{e^-},E_p)}{dE_{e^-}}, \label{Qe+}
\end{equation}
$\mu_{\rm{pp}}=1.45$ is a enhancement factor for collisions
involving heavy nuclei in an SNR (Dermer 1996; Sturner et al. 1997),
$d\sigma(E_{e^-},E_p)/dE_{e^-}$ and $d\sigma(E_{e^+},E_p)/dE_{e^+}$
are the differential cross section for electrons and positrons
produced via p-p interaction, respectively. Since the accelerated
protons colliding with the ambient medium (i.e. p-p interaction) can
produce pions ($\pi^{\pm}$ and $\pi^0$), in which a proton loses
$\sim 1/3$ of its energy per pion-producing collision, there exists
the catastrophic loss for the protons. Sturner et al. (1997) treated
this process as an escape from the system, and the timescale for
this catastrophic loss is
\begin{equation}
\tau_{\rm{pion}}(E_p)=[c\beta_pn_{\rm{SNR}}\sigma_{\rm{pp}}]^{-1}\;\;,
\label{tpp}
\end{equation}
where $\sigma_{\rm{pp}}$ is the inelastic cross section for p-p
interaction, which is (Kelner et al. 2006)
\begin{equation}
\sigma_{\rm{pp}}(E_p)=(34.3+1.88L+0.25L^2)\left[1-\left(\frac{E_{\rm{th}}}{E_p}\right)^4\right]\;\;\mbox{mb},
\label{sigmapp}
\end{equation}
where $L=\ln(E_p/1\mbox{TeV})$, $E_{\rm{th}}= m_p+ 2m_{\pi}+
m^2_{\pi}/2m_p$ is the threshold energy of production of
$\pi$-mesons.

\begin{figure*}
\resizebox{0.8\hsize}{!}{\includegraphics{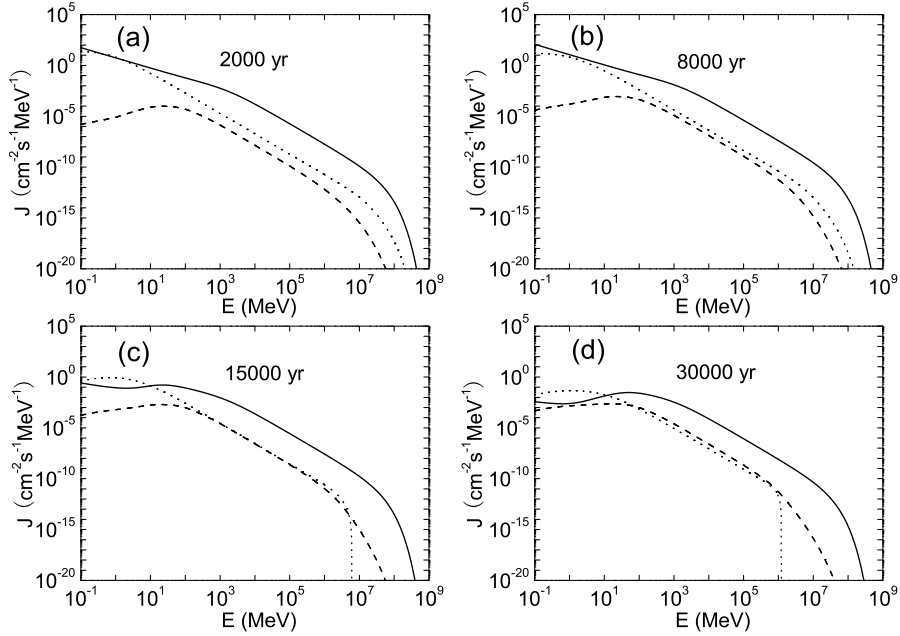}} \caption{The
isotropic intensities of the primary electrons (dotted line), the
primary protons (solid line), and the secondary $e^{\pm}$ pairs
(dashed line) at (a) 2000, (b) 8000, (c) 15000, and (d) 30000 yr,
respectively, for $M_{\rm{ej}}=1.5M_{\odot}$,
$n_{\rm{ISM}}=10\,\rm{cm}^{-3}$, $B_{\rm{ISM}}=5\,\mu\rm{G}$,
$K_{\rm{ep}}=0.001$.} \label{fig:dtinten}
\end{figure*}
\subsection{Photon Production}
We can solve the Eqs. (\ref{Net}) and (\ref{Npt}) to get the volume-
and direction-averaged intensities for electrons and protons at a
given time, and then the multi-waveband non-thermal photon spectra
can be calculated. The non-thermal radiation processes of the
accelerated particles involved in an SNR are synchrotron radiation,
bremsstrahlung, inverse Compton scattering for leptons including
electrons and positrons, and the p-p interaction for protons.

Accelerated electrons can emit photons via synchrotron radiation as
interacting with the ambient magnetic field in an SNR. The
emissivity for synchrotron radiation is given by
\begin{eqnarray}
\nonumber Q_{\rm{syn}}(E_{\gamma},
t)&=&\left(\frac{2\sqrt{3}e^3B_{\rm{SNR}}}{\hbar
E_{\gamma}m_ec^3}\right)\int^{\pi/2}_0d\theta\sin^2\theta\\
&&\times
\int^{E_{e,\rm{max}}}_{E_{e,\rm{min}}}dE_eJ_e(E_e,t)F\left(\frac{E_{\gamma}}{E_{\rm{c}}}\right),
\label{Pusyn}
\end{eqnarray}
where $\theta$ is the electron pitch angle, $E_c=4.2\times
10^6hB_{SNR}\gamma^2_e\sin\theta$, and
\begin{equation}
F(y)=y\int^{\infty}_y dzK_{5/3}(z)\;\;
\end{equation}
with $y=E_{\gamma}/E_{\rm{c}}$, where $K_{5/3}$ is a modified Bessel
function of order 5/3.

For the electron-electron and electron-nucleon bremsstrahlung
radiation, the emissivity is (Sturner et al. 1997)
\begin{eqnarray}
\nonumber Q_{\rm{\rm{brem}}}(E_\gamma, t)&=&4\pi
n_{\rm{SNR}}\Delta^{e,n}_{\rm{\rm{He}}}\\
& & \times\int^{E_{e,\rm{max}}}_{E_{e,\rm{min}}}dE_e
J_e(E_e,t)\left(\frac{d\sigma}{dE_{\gamma}}\right)_{e-e,
p}\;\;\;,\label{Pubrem}
\end{eqnarray}
where $\Delta^{e}_{\rm{He}}$ and $\Delta^{n}_{\rm{He}}$ are
correction factors for the presence of helium
($\Delta^{e}_{\rm{He}}=1.2$ and $\Delta^{n}_{\rm{He}}=1.4$),
$d\sigma_{e-e}/dE_{\gamma}$ and $d\sigma_{e-p}/dE_{\gamma}$
represent the differential electron-electron and electron-nucleon
bremsstrahlung cross sections. Here, we use the approximate formulae
given in Baring et al. (1999).

We use the full Klein-Nishina cross section for the relativistic
electrons to tackle the inverse Compton process, the emissivity can
be expressed as
\begin{eqnarray}
\nonumber Q_{\rm{comp},~j}(E_\gamma,
t)&=&4\pi\int^{\infty}_{0}d\epsilon
n_{\rm{j}}(\epsilon,r)\int^{E_{e,\rm{max}}}_{E_{e,
\rm{thresh}}}dE_e\\
&& \times J_e(E_e,t)F(\epsilon, E_{\gamma}, E_e)\;\;\;,
\end{eqnarray}
where, $\epsilon$ is the target photon energy, $n_{\rm{j}}$ is the
number density of soft photon component $\rm{j}$ with energy density
$U_{\rm{j}}$ and temperature $T_{\rm{j}}$ and is given by
\begin{equation}
n_{\rm{j}}(\epsilon)=\frac {15U_{\rm{j}}}{(\pi k
T_{\rm{j}})^4}\frac{\epsilon^2}{\exp(\epsilon/kT_{\rm{j}})-1}\;\;,
\label{Nj}
\end{equation}
$E_{e,
\rm{thresh}}=[E_{\gamma}+(E^2_{\gamma}+E_{\gamma}(m_ec^2)^2/\epsilon)^{1/2}]/2$
is the lowest energy that electrons can scatter a target photon with
energy $\epsilon$ to energy $E_{\gamma}$. Function $F(\epsilon,
E_{\gamma}, E_e)$ is given by
\begin{eqnarray}
F(\epsilon, E_{\gamma}, E_e)&=&{3\sigma_T\over 4
(E_e/mc^2)^2}{1\over \epsilon} \times\\ & & \left[2q\ln q
+(1+2q)(1-q) \nonumber
 +{(\Gamma q)^2(1-q)\over 2(1+\Gamma q)}\right],
\end{eqnarray}
with $\Gamma=4\epsilon(E_e/mc^2)/mc^2$, $q=E_1/\Gamma (1-E_1)$ with
$E_1=E_{\gamma}/E_e$ and $1/4(E_e/mc^2) < q <1$.

Kamae et al. (2006) presented a series of accurate, convenient
parameterized formulae to calculate the spectra of stable secondary
particles ($\gamma$, $e^{\pm}$, $\nu_e$, $\bar{\nu}_e$, $\nu_{\mu}$,
$\bar{\nu}_{\mu}$) produced in the p-p interactions, which greatly
facilitate calculations involving the p-p interaction in
astronomical environments. The formulae were derived from the
up-to-date p-p interaction model given by Kamae et al. (2005), which
incorporates the logarithmically rising inelastic cross section, the
diffraction dissociation process, and the Feynman scaling violation.
The functional formula reproducing the secondary particle spectra
include the non-diffractive, diffractive, and resonance-excitation
components. For the nondiffractive process, the differential
inclusive cross section to produce a secondary particle is given as
\begin{equation}
\frac{d\sigma_{\rm{ND}}(E_{\rm{sec}})}{d\ln(E_{\rm{sec}})}=
F_{\rm{ND}}(x)F_{\rm{ND},\rm{kl}}(x)\;\;,
\end{equation}
where $E_{\rm{sec}}$ is the energy of the secondary particle,
$x=E_{\rm{sec}}/\rm{GeV}$, $F_{\rm{ND}}(x)$ is the formula
representing the nondiffractive cross section, and
$F_{\rm{ND},\rm{kl}}(x)$ is the formula to approximately enforce the
energy-momentum conservation limits:
\begin{eqnarray}\label{eq:Fnd}
F_{\rm{\rm{ND}}}(x) &=&
a_{0}\exp(-a_{1}(x - a_{3} + a_{2}(x - a_{3})^{2})^{2})  \nonumber\\
&&{}+a_{4}\exp(-a_{5}(x - a_{8} + a_{6}(x - a_{8})^{2} \nonumber\\
&&{}+ a_{7}(x-a_{8})^{3})^{2}),
\end{eqnarray}

\begin{eqnarray}
F_{\rm{ND},\rm{kl}}&=&\frac{1}{\exp\left[W_{\rm{ND,l}}(L_{\rm{min}}-x)\right]+1}\nonumber
\\&&\times
\frac{1}{\exp\left[W_{\rm{ND,h}}(x-L_{\rm{max}})\right]+1}\;\;,
\end{eqnarray}
where $L_{\rm{min}}$ and $L_{\rm{max}}$ are the lower and upper
kinematic limits imposed and $W_{\rm{ND,l}}$ and $W_{\rm{ND,h}}$ are
the widths of the kinematic cutoffs, $L_{\rm{min}}=-2.6$ for all
secondary particles, and the other parameters are given in Table
\ref{Lmax} (Kamae et al. 2006).
\begin{center}
\begin{table}
\caption{Kinematic limit parameters for the nondiffractive process
(Kamae et al. 2006).} \label{Lmax}
\begin{center}
\begin{tabular}{cccc}
\hline\hline
  & $L_{\rm{max}}$ & $W_{\rm{\rm{ND},l}}$ & $W_{\rm{\rm{ND},h}}$ \\
\hline
$\gamma$ & 0.96$\log_{10}(T_{p})$ & 15 & 44 \\
$e^{-}$ & 0.96$\log_{10}(T_{p})$ & 20 & 45 \\
$e^{+}$ & 0.94$\log_{10}(T_{p})$ & 15 & 47 \\
$\nu_{e}$ & 0.98$\log_{10}(T_{p})$ & 15 & 42 \\
$\bar{\nu}_{e}$ & 0.98$\log_{10}(T_{p})$ & 15 & 40 \\
$\nu_{\mu}$ & 0.94$\log_{10}(T_{p})$ & 20 & 45 \\
$\bar{\nu}_{\mu}$ & 0.98$\log_{10}(T_{p})$ & 15 & 40 \\
\hline
\end{tabular}\end{center}
\end{table}
\end{center}
The differential inclusive cross section to produce a secondary
particle for the diffractive process can be represented as
\begin{equation}
\frac{d\sigma_{\rm{diff}}(E_{\rm{sec}})}{d\ln(E_{\rm{sec}})}=
F_{\rm{diff}}(x)F_{\rm{kl}}(x)\;\;,
\end{equation}
where $F_{\rm{diff}}(x)$ represents the diffractive cross section,
and $F_{\rm{kl}}(x)$ is a function to enforce the energy-momentum
conservation:
\begin{eqnarray}\label{eq:Fdiff}
F_{\rm{diff}}(x) &=& b_{0}\exp(-b_{1}((x - b_{2})/(1 + b_{3}(x - b_{2})))^{2})+  \nonumber\\
&&b_{4}\exp(-b_{5}((x - b_{6})/(1 + b_{7}(x - b_{6})))^{2}),
\end{eqnarray}
\begin{equation}\label{eq:Fdiffkl}
F_{\rm{kl}}(x) = \frac{1}{\exp{(W_{\rm{\rm{diff}}}(x -
L_{\rm{max}}))} + 1},
\end{equation}
here $W_{\rm{\rm{diff}}}=75$,
$L_{\rm{max}}=\log_{10}(T_{p}/\rm{GeV})$.

For the resonance-excitation processes, the function is
\begin{equation}\label{eq:delta}
\frac{d\sigma_{\rm{res}}(E_{\rm{sec}})}{d
\ln(E_{\rm{sec}})}=F_{\rm{res}}(x)F_{\rm{kl}}(x),
\end{equation}
$F_{\rm{res}}(x)$ represents the cross section, and
$F_{\rm{kl}}(x)$, which is same as that for the diffraction process,
enforces the energy-momentum conservation:
\begin{eqnarray}\label{eq:Fdelta}
\nonumber F_{\rm{res}}(x) &=&c_{0}\exp(-c_{1}((x - c_{2})/(1 +
c_{3}(x - c_{2}) \\
& &{}+ c_{4}(x - c_{2})^{2}))^{2}).
\end{eqnarray}

A renormalization factor, $r(T_{p})$, must be multiplied to the
final spectrum to ensure that the parameterized model reproduces the
experimental $\pi^0$ multiplicity after the readjustment in the
resonance-excitation region of $T_{p}$. The exact expressions for
$r(T_{p})$ and other parameters involved in the formulae are shown
in Kamae et al. (2006).  Finally, the differential cross section for
the secondary particles can be expressed as
\begin{eqnarray}\nonumber
\frac{d\sigma(E_{\rm{sec}}, E_p)}{dE_{\rm{sec}}} &=&
\frac{d\sigma_{\rm{ND}}(E_{\rm{sec}}, E_p)}{dE_{\rm{sec}}} +
\frac{d\sigma_{\rm{diff}}(E_{\rm{sec}}, E_p)}{dE_{\rm{sec}}} \\
\nonumber && {}+ \frac{d\sigma_{\rm{\Delta(1232)}}(E_{\rm{sec}},
E_p)}{dE_{\rm{sec}}} \\&& {}+
\frac{d\sigma_{\rm{res(1600)}}(E_{\rm{sec}},
E_p)}{dE_{\rm{sec}}}\;\; .
\end{eqnarray}

The resulting particle intensities calculated with the model are
shown in Fig. \ref{fig:dtinten}. The parameters we choose are
$M_{\rm{ej}}=1.5M_{\odot}$, $n_{\rm{ISM}}=10\rm{cm}^{-3}$,
$B_{\rm{ISM}}=5\mu\rm{G}$, $K_{\rm{ep}}=0.001$, for ages of 2000,
8000, 15000, 30000 yr. For these parameters, the primary particle
sources turn off at $t=t_{\rm{rad}}=1.01\times10^{4}$ yr.
 The turnover at the low energy
region for both electrons and protons is due to Coulomb loss, and
the high-energy turnover for electrons is due to synchrotron loss.
More importantly, the primary electrons dominate the secondary
$e^{\pm}$ pairs for a young age, however, the secondary $e^{\pm}$
pairs become more and more important and eventually surpass the
primary electrons as the age of the SNR increases, the reason is
that the sources of the primary particles are cut off in the
radiative phase of the SNR and the primary electrons experience
strong synchrotron energy loss, however, the energy loss for the
protons is negligible due to the small cross section for the p-p
interaction, and then the secondary particles can be produced
continuously for a long time.
\begin{figure*}
\resizebox{\hsize}{!}{\includegraphics{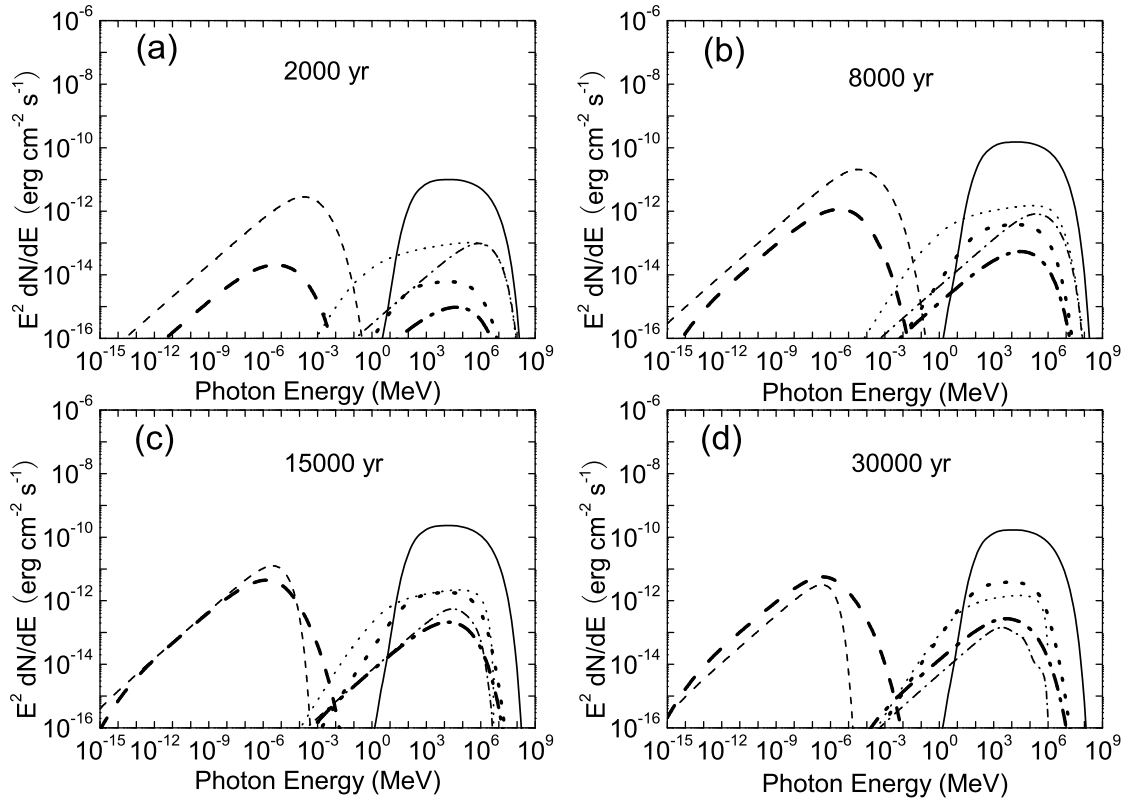}} \caption{The
resulting calculated multi-wavelength spectra at (a) 2000, (b) 8000,
(c) 15000, and (d) 30000 yr, respectively. The distance is set to 2
kpc and the other parameters are same as those in Fig.
\ref{fig:dtinten}. The emission due to synchrotron (dashed line),
bremsstrahlung (dotted line), Compton scattering (dash-dotted line)
of the primary electrons (thin) and the secondary $e^{\pm}$ pairs
(thick), and the p-p interaction (solid line) are shown in each
panel.} \label{fig:dtpu}
\end{figure*}

We show the resulting multi-wavelength non-thermal spectra in Fig.
\ref{fig:dtpu}. The distance to the SNR is chosen to be 2 kpc, and
other parameters are same as those in Fig. \ref{fig:dtinten}. It is
obvious that the emission from radio to X-ray band for the SNR comes
from the synchrotron radiation of leptons, and the high-energy
photons can from bremsstrahlung and inverse Compton scattering of
leptons, and meson decay due to the hadronic interactions. More
importantly, (1) the roll-off energy of the synchrotron radiation of
the primary electrons deceases quickly when the SNR ages in the
radiative phase, however, the synchrotron emission of the secondary
$e^{\pm}$ pairs become more and more important and eventually
surpass the synchrotron radiation from the primary electrons when
the SNR ages due to the secondary $e^{\pm}$ pairs can be produced
for a long time via the p-p interaction; (2) most VHE $\gamma$-rays
from the SNR are of hadronic origin with those parameters, and the
hadronic emission from an SNR is similar during a long time when the
SNR is in the radiative phase because the energy loss rate for the
contained protons is small. Of course, the photons with energy
smaller than 10 MeV from some old SNRs mainly come from the
radiation of the secondary $e^{\pm}$ pairs.

\begin{figure*}
\resizebox{0.8\hsize}{!}{\includegraphics{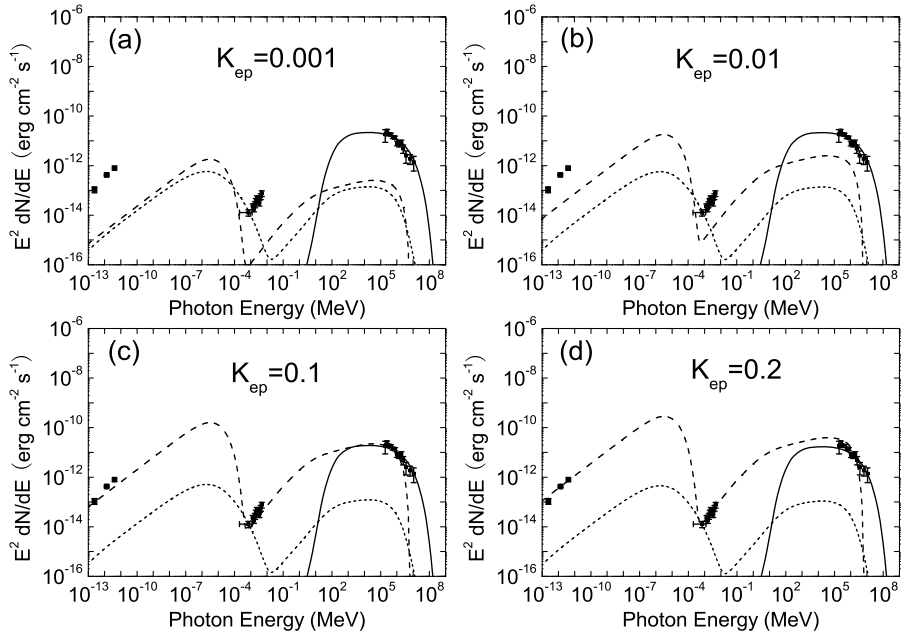}}
\caption{Comparisons of the model results with the multi-band
observations for the SNR G8.7$-$0.1. The radiation from the primary
electrons (dashed line), the primary protons (solid line), and the
secondary $e^{\pm}$ pairs (dotted line) with data points from
Odegard (1986), Kassim \& Weiler (1990), Green (2006) (radio), Cui
\& Konopelko (2006) (X-ray), Aharonian et al. (2006b)
(\textit{HESS}) are indicated in each panel. The parameters are a
distance of 6\,kpc, an age of 15000 yr,
$M_{\rm{ej}}=1.5\,M_{\odot}$, $n_{\rm{ISM}}=8\, \rm{cm}^{-3}$,
$B_{\rm{ISM}}=6\,\mu\rm{G}$, (a) $K_{\rm{ep}}=0.001$, (b)
$K_{\rm{ep}}=0.01$, (c) $K_{\rm{ep}}=0.1$, (d) $K_{\rm{ep}}=0.2$.}
\label{fig:dk216}
\end{figure*}

\begin{figure*}
\resizebox{0.8\hsize}{!}{\includegraphics{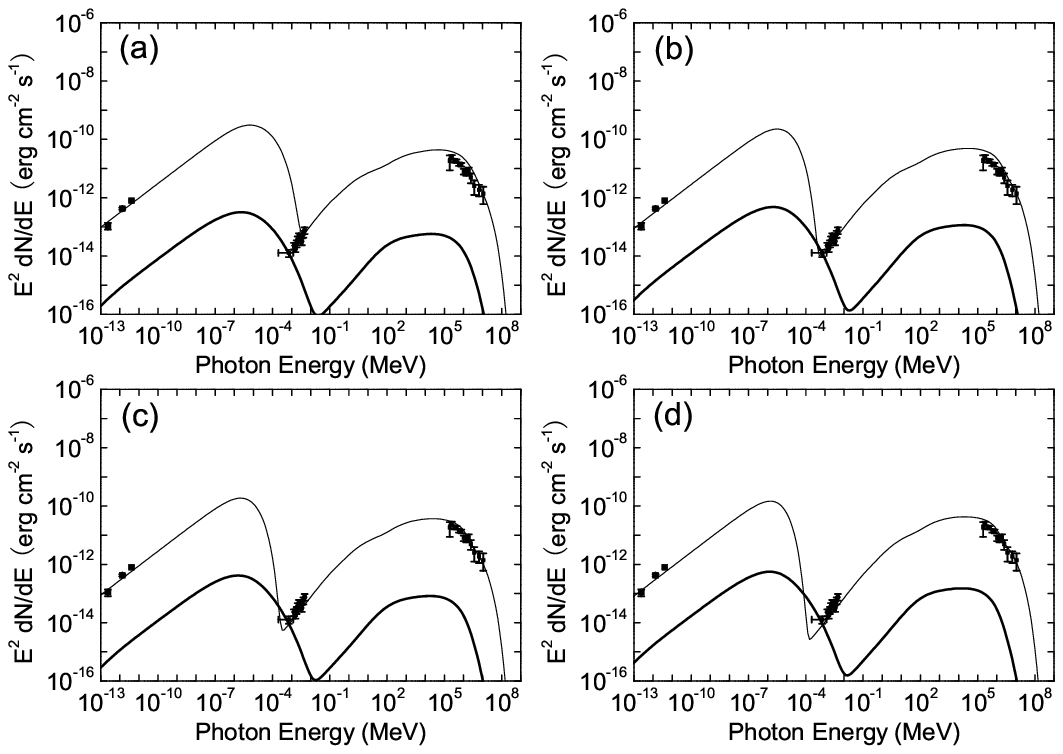}} \caption{
Comparisons of the model spectra  with the observations for the SNR
G8.7$-$0.1 with $K_{\rm{ep}}=0.15$, (a) $n_{\rm{ISM}}=6\,
\rm{cm}^{-3}$, an age of 15000 yr; (b) $n_{\rm{ISM}}=8\,
\rm{cm}^{-3}$, an age of 15000 yr; (c) $n_{\rm{ISM}}=6\,
\rm{cm}^{-3}$, an age of 18000 yr; (b) $n_{\rm{ISM}}=8\,
\rm{cm}^{-3}$, an age of 18000 yr. The whole emission from the
primary particles and the secondary $e^{\pm}$ pairs are indicated by
thin solid line and thick solid line respectively. Others are same
as those in Fig. \ref{fig:dk216}.} \label{fig:dn216}
\end{figure*}

\section{Application to old SNRs} \label{sec:application}

\subsection{SNR G8.7$-$0.1}
Aharonian et al. (2006b) reported a survey of the inner part of the
Galactic Plane in very high energy $\gamma$-rays observed with
\textit{HESS} Cherekov telescope system, and fourteen previously
unknown sources were detected at a significance level greater than
$4\sigma$ after accounting for all trials involved in the search. A
source detected in the survey is HESS~J1804$-$216, which is the
brightest one of the new sources, with a steep photon index of
$2.72\pm0.12$ and a flux of nearly $25\%$ of the flux from the Crab
Nebula above 200 GeV. The source can be associated with the
south-western part of the shell of the SNR G8.7$-$0.1 of radius 26
arc minutes with flux 80 Jy at 1 GHz (Green 2006), or the young
Vela-like pulsar PSR J1803$-$2137 with a spin-down age of 16000 yr.
Fatuzzo et al. (2006) presented a simple model for the TeV emission
from HESS~J1804$-$216, they found that the SNR G8.7$-$0.1 can
plausibly account for all the known radiative characteristics of
HESS~J1804$-$216, and therefore the SNR G8.7$-$0.1 was probably the
source of the TeV photons originating from the direction of
HESS~J1804$-$216.

By associating G8.7$-$0.1 with coincident HII regions, Kassim \&
Weiler (1990) estimated a distance to the SNR of $6\pm1$ kpc, with
angular size of $\sim50$ arcmin, a physical size of $\sim80$pc, and
an age of 15000 yr. Cui \& Konopelko (2006) reported the
high-resolution X-ray observations taken with the Chandra X-Ray
Observatory of the field that contains the TeV $\gamma$-ray source
HESS~J1804$-$216, and a total of 11 discrete sources were detected.
Among those sources, only one, CXOU~J180351.4$-$213707, is the most
probable X-ray counterpart of HESS~J1804$-$216, is significantly
extended with photon index $1.2^{+0.5}_{-0.4}$.

We now assume that the sources CXOU J180351.4 $-$ 213707 and HESS
J1804$-$216 associate with the SNR G8.7$-$0.1. To make the resulting
spectrum consistent with the observations by \textit{Chandra} with a
low energy flux and a photon index about 1.2, the SNR must be in the
radiative phase when the roll-off energy of the synchrotron
radiation of the primary electrons decreases quickly and the X-rays
detected by Chandra X-Ray Observatory are most probably from the
bremsstrahlung radiation or inverse Compton scattering of leptons,
then $n_{\rm{ISM}}$ should be greater than $\sim 5$ cm$^{-3}$ for an
age of 15000 yr.

We now describe the detailed process in modeling the multi-waveband
spectrum for the SNR. First of all, we investigate the influence of
$K_{\rm{ep}}$ on the final spectra with parameters a distance of 6
kpc, an age of 15000 yr, $M_{\rm{ej}}=1.5\,M_{\odot}$,
$B_{\rm{ISM}}=6\,\mu\rm{G}$, and $n_{\rm{ISM}}=8\, \rm{cm}^{-3}$ for
the SNR G8.7$-$0.1. The calculated results are shown in Fig.
\ref{fig:dk216}. The spectrum in the X-ray band depends heavily on
$K_{\rm{ep}}$, and the value of $K_{\rm{ep}}$ must be around 0.1 in
order to match the observed data in X-ray band for the SNR.

\begin{figure}
\resizebox{\hsize}{!}{\includegraphics{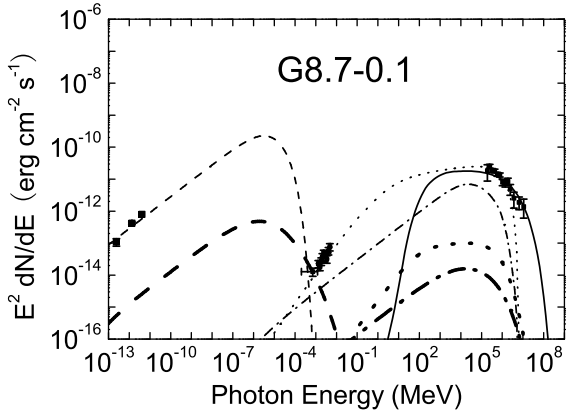}} \caption{The
comparison of detailed spectra calculated by the model with the
observations for the SNR G8.7$-$0.1. The parameters are same as
those for panel (c) in Fig. \ref{fig:dn216}. Synchrotron (dashed
line), bremsstrahlung (dotted line), Compton (dash-dotted line) of
the primary electrons (thin) and the secondary $e^{\pm}$ pairs
(thick), meson decay due to the p-p interaction (black solid line)
are indicated in the figure. } \label{fig:216}
\end{figure}

Fig. \ref{fig:dn216} shows the resulting spectra from the SNR for
different $n_{\rm{ISM}}$ and ages. For an age of 15000 yr, the
roll-off energy of the synchrotron emission is still high for
$n_{\rm{ISM}}=6\, \rm{cm}^{-3}$, which conflicts with the
\textit{Chandra} observation, and the level of the TeV emission is
greater than the observed one by \textit{HESS} both for
$n_{\rm{ISM}}=6\, \rm{cm}^{-3}$ and for $n_{\rm{ISM}}=8\,
\rm{cm}^{-3}$ (see penal (a) and penal (b) in Fig. \ref{fig:dn216}).
The model results can fit the multi-band observations well with a
lager age of 18000 yr, moreover, the calculated spectrum in the band
0.1 -- 1 TeV for $n_{\rm{ISM}}=6\, \rm{cm}^{-3}$ fit the TeV
observations better than that for $n_{\rm{ISM}}=8\, \rm{cm}^{-3}$
(see penal (c) and penal (d) in Fig. \ref{fig:dn216}). Furthermore,
the contributions to the final spectrum of different radiation
components with the same parameters as those in penal (c) of Fig.
\ref{fig:dn216} are shown in Fig.\ref{fig:216}. Obviously from
Fig.\ref{fig:216}, for the SNR, (1) the emission from the primary
electrons dominates that from the secondary $e^{\pm}$ pairs in the
entire energy band except in the narrow soft X-ray band around 0.5
keV; (2) the detected radio emission is mainly from the synchrotron
radiation of the primary electrons whereas the X-rays observed with
\textit{Chandra} are primarily produced via the bremsstrahlung of
these electrons; (3) the TeV photons with energies $<1$ TeV are
primarily from both the bremsstrahlung of the primary electrons and
the p-p interaction of the primary protons, however, those with
higher energies are mainly from the p-p interactions due to the
high-energy protons collide with the ambient matter.

\begin{figure*}
\resizebox{0.8\hsize}{!}{\includegraphics{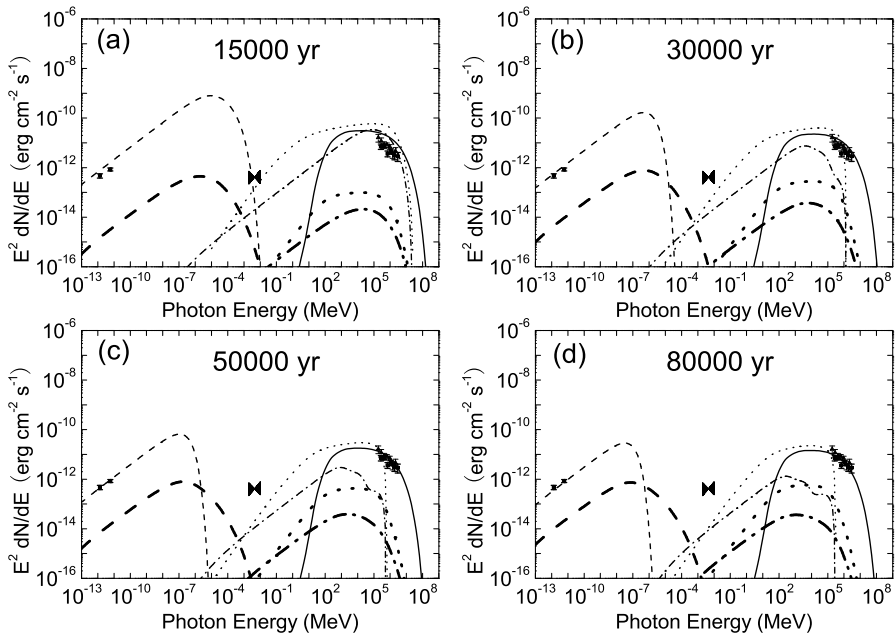}} \caption{
Comparisons of the calculated spectra with the observations for the
SNR G23.3$-$0.3 at different ages. Synchrotron (dashed line),
bremsstrahlung (dotted line), Compton (dash-dotted line) of the
primary electrons (thin) and the secondary $e^{\pm}$ pairs (thick),
meson decay due to the p-p interaction (black solid line) with
radio, X-ray (Tian et al. 2007), and VHE $\gamma$-ray with
\textit{HESS} (Aharonian et al. 2006b) and \textit{MAGIC} (Albert et
al. 2006) observational results are indicated in the figure. The
other parameters are $M_{\rm{ej}}=1.5\,M_{\odot}$,
$n_{\rm{ISM}}=6\,\rm{cm}^{-3}$, $B_{\rm{ISM}}=5\,\mu\rm{G}$,
$K_{\rm{ep}}=0.2$, a distance of 4\,kpc, and an age of (a) 15000 yr,
(b) 30000 yr, (c) 50000 yr, (d) 80000 yr.} \label{fig:dy087}
\end{figure*}

\subsection{SNR G23.3$-$0.3}

HESS~J1834$-$087 is another extended TeV source recently discovered
in the \textit{HESS} survey of the inner Galaxy in VHE
$\gamma$-rays. The source, which is spatially coincident with the
SNR G23.3$-$0.3 (W41), has a size of 12 arc minutes and a
$\gamma$-ray flux of 8\% of the flux from the Crab Nebula above 200
GeV, with a photon index of $2.45\pm0.16$ (Aharonian et al. 2006b).
Moreover, Albert et al. (2006) presented the observations of
HESS~J1834$-$087 with the Major Atmospheric Gamma Imaging Cherenkov
telescope (\textit{MAGIC}), and the differential $\gamma$-ray flux
can be expressed as $dN/dAdtdE =
(3.7\pm0.6)\times10^{-12}(E/\rm{TeV})^{-2.5\pm0.2}\;\rm{cm}^{-2}\;\rm{s}^{-1}\;\rm{TeV}^{-1}$.

G23.3$-$0.3 is an asymmetric shell-type SNR with radio and X-ray photons having been detected.
 In radio band, the SNR has a spectral
index of 0.5 and a flux of 70 Jy at 1 GHz (Green 2006). Moreover, Tian et al.
(2007) reported new H I observations from the VLA Galactic Plane
System and a new \xmm\ observation for HESS~J1834$-$087. They
concluded that G23.3$-$0.3 is an old SNR with a distance of
$4\pm0.2$ kpc and an age of $\sim10^5$ yr. Furthermore,
 a density of
6 cm$^{-3}$ was estimated around the SNR from the VGPs H I line emission associated with the SNR (Tian et al. 2007).
 The new \xmm\ observation
indicated diffuse X-ray emission within the \textit{HESS} source and
suggested an association between the X-ray and the VHE $\gamma$-ray
emission. The intrinsic X-ray spectrum from 2 to 10 keV could be
characterized as a heavily absorbed power law with a flux of
$7.0\times10^{-13}\;\rm{erg}\;\rm{s}^{-1}\;\rm{cm}^{-2}$, and a
photon index of $2.0^{+0.7}_{-0.8}$ (see details in Tian et al.
2007).

\begin{figure}
\resizebox{\hsize}{!}{\includegraphics{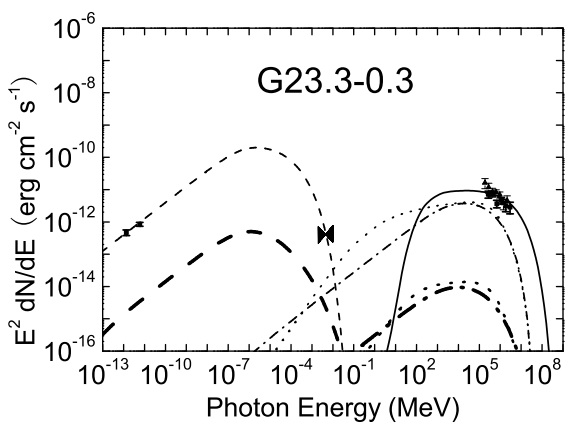}} \caption{The
comparison of detailed spectra calculated by the model with the
observations for the SNR G23.3$-$0.3 with an age of 29800 yr.
 The other parameters are
$M_{\rm{ej}}=1.5\,M_{\odot}$, $n_{\rm{ISM}}=1.5\,\rm{cm}^{-3}$,
$B_{\rm{ISM}}=8\,\mu\rm{G}$, $K_{\rm{ep}}=0.05$, and a distance of
4\,kpc. Others are same as those in Fig. \ref{fig:dy087}.}
\label{fig:087}
\end{figure}

We now apply the model in this paper to the old SNR G23.3$-$0.3,
which has been observed in radio, X-ray and VHE $\gamma$-ray bands.
Firstly, using the similar process as that for the SNR G8.7$-$0.1,
we find $K_{\rm{ep}}$ must be about 0.2 for the SNR G23.3$-$0.3 with
parameters, a distance of 4 kpc, $M_{\rm{ej}}=1.5\,M_{\odot}$,
$n_{\rm{ISM}}=6\,\rm{cm}^{-3}$, $B_{\rm{ISM}}=5\,\mu\rm{G}$, and an
age of several $10^4$ yr. Fig. \ref{fig:dy087} shows the resulting
spectra for different ages. For these parameters, the radiative age
is $1.43\times10^4$ yr and then the emission in each panel of Fig.
\ref{fig:dy087} is in the radiative phase. The observations in X-ray
band can be explained as the synchrotron and bremsstrahlung
radiation of the primary electrons for an age of 15000 yr, however,
the calculated results are significantly high above the VHE
$\gamma$-ray observations by \textit{HESS} and \textit{MAGIC} (see
panel (a) in Fig. \ref{fig:dy087}). When the radiative SNR ages, the
cut-off energy of the primary electrons deceases quickly and the
high-energy protons undergo both catastrophic energy losses due to
the p-p collisions and energy-dependent diffusion. As a result, the
roll-off energy of the synchrotron radiation of the primary
electrons decreases quickly and the number of the TeV photons from
the SNR reduces too. We can see from panel (d) in Fig.
\ref{fig:dy087} that the radio and VHE $\gamma$-ray observations can
be explained by the model with an age of 80000 yr, however, the
observed flux in the X-ray band can not be reproduced. To make the
model results consistent with the flux of the observations in the
X-ray band, the age of the SNR should not be much bigger than the
radiative age of the SNR. Furthermore, a smaller $n_{\rm{ISM}}$
should be chosen to make the calculation spectrum in the VHE
$\gamma$-ray band consistent with the \textit{HESS} and
\textit{MAGIC} observations.

Finally, $n_{\rm{ISM}}$ is chosen to be 1.5 cm$^{-3}$ with other
parameters an age of $2.98\times10^4$ yr,
$M_{\rm{ej}}=1.5\,M_{\odot}$, $B_{\rm{ISM}}=8\,\mu\rm{G}$,
$K_{\rm{ep}}=0.05$ to model the multi-band non-thermal emission from
the SNR G23.3$-$0.3 and the resulting spectra are shown in Fig.
\ref{fig:087}. The radiative age for these parameters is
$2.97\times10^4$ yr and the roll-off energy of the synchrotron
radiation of the primary electrons is not reduced mostly with an age
of $2.98\times10^4$ yr. It is obvious from Fig. \ref{fig:087} that
(1) the emission from the primary electrons dominates that from the
secondary $e^{\pm}$ pairs in all bands and the emission from radio
and X-ray band is mainly from the synchrotron radiation of the
primary electrons; (2) the TeV photons are produced mainly via the
p-p interaction.

\section{Conclusions and Discussions} \label{sec:discussion}
In this paper, by taking into account the evolution of secondary
$e^{\pm}$ pairs produced via the p-p interaction due to high-energy
protons collide with the ambient matter in an SNR, we developed a
time-dependant model of multi-waveband non-thermal particle and
photon spectra both for young and for old shell-type SNRs. The
primary electrons accelerated directly by the shock wave usually
dominate the secondary $e^{\pm}$ pairs produced from the p-p
collisions for a young SNR, however, the emission from the secondary
$e^{\pm}$ pairs becomes more and more important and eventually
surpasses the radiation from the primary electrons as the SNR grows
old because the source of the primary electrons is cut off and the
synchrotron radiation loss is significant for electrons when the SNR
is in the radiative phase, whereas the secondary $e^{\pm}$ pairs can
be produced for a long time due to the small cross section for the
p-p interaction (see Fig. \ref{fig:dtpu}). $K_{\rm{ep}}$, which can
not be directly deduced from the observations now and is usually
limited by comparison of the model results with the multi-band
observations for a given SNR, is also a crucial parameter to
determine whether the contribution to the final non-thermal emission
of the secondary $e^{\pm}$ pairs dominates that of the primary
electrons. For an old SNR with small $K_{\rm{ep}}$, the non-thermal
photons with energies $<10\,\rm{MeV}$ are usually produced mainly
via the radiations of the secondary $e^{\pm}$ pairs (see Fig.
\ref{fig:dtpu}).

The model is applied to two old shell-type SNRs, G8.7$-$0.1 and
G23.3$-$0.3, whose VHE $\gamma$-rays are detected in the
\textit{HESS} survey of the inner Galaxy. G8.7$-$0.1 is a shell-type
SNR with an age of $\sim15000$ yr. The SNR must be in the radiative
phase to make the calculated results consistent with the
\textit{Chandra} X-ray observations. Finally, a set of parameters, a
distance of 6 kpc, an age of 18000 yr, $M_{\rm{ej}}=1.5\,M_{\odot}$,
$n_{\rm{ISM}}=6\,\rm{cm}^{-3}$, $K_{\rm{ep}}=0.15$, and
$B_{\rm{ISM}}=6\,\mu\rm{G}$ is used to model the multi-wavelength
non-thermal emission from the SNR G8.7$-$0.1. It can be concluded
that (1) the radio emission from the SNR is mainly from the
synchrotron radiation of the primary electrons, whereas the
non-thermal photons with energies around 0.1 keV primarily come from
the synchrotron radiation of the secondary $e^{\pm}$ pairs produced
from the p-p collisions; (2) the observed X-ray photons with
energies ranging from 2 to 10 keV are produced mainly via the
bremsstrahlung of the primary electrons; (3) the TeV photons with
energies $<1$ TeV are primarily from both the bremsstrahlung of the
primary electrons and the p-p collisions of the primary protons,
however, those with higher energies are produced mainly through the
p-p interaction (see Fig. \ref{fig:216}). Another old SNR is
G23.3$-$0.3, which is an asymmetric shell-type SNR and spatially
coincide with HESS~J1834$-$087. Tian et al. (2007) reported new H I
observations from the VLA Galactic Plane System and a new \xmm\
observation for HESS~J1834$-$087, and argued that G23.3$-$0.3 is an
old SNR with a distance of $4\pm0.2$ kpc and an age of $\sim10^5$
yr, moreover, a density of 6 cm$^{-3}$ was estimated around the SNR
from the VGPs H I line emission associated with the SNR. We find
that, using $n_{\rm{ISM}}=6\,\rm{cm}^{-3}$ and an age of $\sim10^5$
yr with other appropriate parameters, the model result agrees with
the radio and VHE $\gamma$-ray observations, but the calculated
X-ray flux is significantly lower than the observed one given by
\xmm\ (see Fig. \ref{fig:dy087}). A smaller age and a smaller
$n_{\rm{ISM}}$ are needed to make the calculation result consistent
with the multi-band observations. Finally, we use parameters a
distance of 4 kpc, $M_{\rm{ej}}=1.5\,M_{\odot}$, an age of
$2.98\times10^4$ yr, $n_{\rm{ISM}}=1.5\,\rm{cm}^{-3}$,
$B_{\rm{ISM}}=8\,\mu\rm{G}$, and $K_{\rm{ep}}=0.05$ to model the
multi-band non-thermal emission from the SNR G23.3$-$0.3. The
modeling results can be consistent with both the radio, VHE
$\gamma$-ray observations and the X-ray flux with these parameters.
Moreover, the emission from the secondary $e^{\pm}$ pairs is
negligible and the multi-wavelength photons with those parameters
for the SNR mainly come from the radiation of the primary particles,
and the TeV photons primarily have hadronic origin (see Fig.
\ref{fig:087}).

The model in this paper includes the contribution of the secondary
$e^{\pm}$ pairs produced from the p-p collisions to the non-thermal
emission from SNRs. The secondary $e^{\pm}$ pairs can be produced
for a long time and experience energy losses and diffusion, and the
contribution of the secondary $e^{\pm}$ pairs to the final
non-thermal emission from an SNR can become more and more important
and eventually surpasses that of the primary electrons when the SNR
ages in the radiative phase. Whether the contribution of the
secondary $e^{\pm}$ pairs to the final non-thermal emission is
prominent or not depends heavily on the SNR age, $K_{\rm{ep}}$ and
$n_{\rm{ISM}}$. These parameters can be limited by the comparison of
the model result with the multi-wavelength observations for a given
SNR, however, the shortage of precise observations and uncertainties
such as the initial eject mass and the initial velocity of the shock
lessen the reliability of the calculated results. On the other hand,
for an old SNR with observed TeV emission, if the TeV photons
 primarily have hadronic origin, the emission from the secondary $e^{\pm}$
pairs is usually prominent. The morphology of the resulting spectrum
for an SNR in the energy range between 10 MeV to 1 TeV directly
relates to the primary origin of the VHE photons. Fortunately, the
Gamma Ray Large Area Space Telescope (\textit{GLAST}) have a good
sensitivity and resolution in the energy range between 30 MeV and
300 GeV, and the observations with \textit{GLAST} will give a direct
evidence on whether the TeV photons from an SNR mainly have hadronic
origin or leptonic origin. Obviously, \textit{GLAST} will give more
limits on the parameters in the modeling process with the model in
this paper for an SNR, and then we can get more insights on the
acceleration and photon production processes involved in SNRs.

\section*{acknowledgements}
We are grateful to H. Bartko for providing the \textit{MAGIC} data
points and an anonymous referee for his/her very constructive
comments. This work is partially supported by a Distinguished Young
Scientists grant from the National Natural Science Foundation of
China (NSFC 10425314), NSFC grant 10463002, NSFC grant 10778702, and
grant from the Department of Education of Yunnan Province
(07J51074). We appreciate the use of the high performance computer
facilities at Yunnan University, Kunming, PRC.


\label{lastpage}

\end{document}